# Representation Learning based and Interpretable Reactor System Diagnosis Using Denoising Padded Autoencoder

Chengyuan Li[1], Zhifang Qiu[1], Zhangrui Yan[2], Meifu Li[1, *]

*Corresponding author. Email: meifu_lee@163.com. Tel: (+86)13880702386

[1]*Science and Technology on Reactor System Design Technology Laboratory, Nuclear Power Institute of China, Chengdu, 610213, China*

[2]*CNNC Key Laboratory on Nuclear Reactor Thermal Hydraulics Technology, Nuclear Power Institute of China, Chengdu, 610213, China*

## Abstract

With the mass construction of Gen III nuclear reactors, it is a popular trend to use deep learning (DL) techniques for fast and effective diagnosis of possible anomalies to ensure the safety of reactor systems under accidents. To improve the robustness of diagnostic model to crippled and noisy monitoring data, and to enhance the interpretability of the diagnostic outputs, this paper focuses on three aspects. First, a novel Denoising Padded Autoencoder (DPAE) model is proposed for representation extraction of monitoring data, and such representation extractor is effective on disturbed data with signal-to-noise ratios up to 25.0 and monitoring data missing up to 40.0%. Secondly, a diagnostic paradigm using DPAE encoder followed by shallow statistical learning algorithms is proposed, and such stepwise diagnostic approach is tested on disturbed datasets with 41.8% and 80.8% higher classification and regression task evaluation metrics, in comparison with the end-to-end diagnostic approaches. Finally, a hierarchical interpretation algorithm using SHAP and feature ablation is presented to analyze the importance of the input monitoring parameters and validate the effectiveness of the high importance parameters. The results of this study inform the development of robust, interpretable, intelligent reactor anomaly diagnosis systems in situations with high safety requirements.

***Keywords***: *Nuclear Reactor; Intelligent Fault Diagnosis; Representation Learning; Explainable Artificial Intelligence*

Code for DPAE is available at

https://github.com/lichengyuan98/deep-learning-elements/blob/master/my_models/Padded_AE.py

## 1. Introduction

### 1.1 Background

In worldwide "carbon neutrality" and to reduce greenhouse gas emissions from energy production and achieve the goal of limiting global average warming to 2°C by 2050, pressurized water nuclear reactors are an indispensable option for addressing global climate change and promoting energy recovery in the post-epidemic era[1].

Current nuclear power development is gradually emerging from the adverse effects of the 2011 Fukushima nuclear accident. As of July 2022, the world's installed nuclear power capacity has reached 390 GWe and contributes 2,553 TWh of electricity in 2020, generating 10% of the world's electricity and accounting for 28% of low-carbon energy. In FY 2021, the OECD's International Energy Agency (IEA) released its World Energy Outlook report. The report forecasts that globally installed nuclear power capacity will grow by 26% between 2020 and 2050 to 525 GWe, based on established national policies[2].

The construction of Gen III nuclear power projects is booming globally, and therefore requires special attention to safety issues in their operation. Nuclear power capacity is steadily growing worldwide, with 51 nuclear power projects under construction as of July 2022, of which Gen III pressurized water reactor account for 84.3%[3]. Although Gen III nuclear power plants have reduced the probability of serious accidents by two orders of magnitude compared to Gen II plants, the possibility of serious accidents is still not eliminated. Thus, safety and reliability of Gen III nuclear power plants, especially under abnormal transients, should be enhanced with efficiency.

**1.2 Problem definition**

To safeguard the health status of the reactor during operation, the principle of prognostics and health management (PHM) is commonly used. PHM comprises five components, namely data acquisition, monitoring and anomaly detection, fault diagnosis, prognostics, and planning and decision making [4]. This framework provides early warning for reactor equipment degradation and initiation of events through detection, diagnosis, and prognosis methods.

When abnormal transients occur, proper diagnosis of the cause and severity is a critical step, as incorrect diagnosis will increase the risk of core damage. Fast, accurate and robust diagnosis is a challenging task due to the complexity of the system, the highly non-linear correlation of the monitoring parameters, and the uncertainty of the monitoring devices due to disturbances.

To assist the operator in the identification of initiating events, a series of intelligent fault diagnosis (IFD) methods have been developed. The goal of IFD is to automatically uncover the relationship between the monitored signals and the system health status, which is one of the important aspects of PHM and is the basis for the subsequent treatments. According to one classification, these diagnostic methods can be broadly categorized into three groups, namely model-based, expert knowledge-based, and data-based diagnostics[5]. Model-based methods perform quantitative analysis through filtering, parity and frequency transformation, and qualitative analysis through graphical models and qualitative physics, but the results are preliminary and need to be used with other methods. Expert knowledge-based methods, also known as rule-based methods, are used to determine the relationship between symptoms and causes in an "if-then" format, but the problem lies in the management of a large knowledge base and the failure that occurs in the absence of rules. Data-based methods rely on a large amount of historical process data to develop diagnostic models, and due to the increase in computer computing power in recent decades, their diagnostic models are more scalable, faster, and more accurate than the other two types of methods, so they have gained more attention from researchers.

## 1.3 Related works

Previous efforts to develop diagnostic methods for IFD have taken statistical learning based[6–8], Bayesian network based[9–11] and Deep Learning (DL) based algorithms[12–16].DL based methods have a powerful domain adaptation capability due to their structural characteristics of multilayer nonlinear mapping and the use of back-propagation gradient descent optimization algorithms.

For the task of diagnosing reactor accidents, DL techniques can often be used as two instruments, an identifier for the type and severity of anomalies, and a feature extractor for anomalous transients, respectively. In terms of using DL techniques as identifiers, an accident diagnosis model based on LSTM and CNN has been proposed by Saeed et al., which can not only identify the 'unknown' transient type, but also calculate the severity of the accident[17]. Wang et al. used convolution layer instead of full connection layer in traditional GRU neural network, whose hyper parameters are searched by particle swarm optimization algorithm, which achieved good results in experiments[18]. In case of using DL techniques as feature extractors, Li et al. proposed a method for feature extraction and clustering of transients, using CNN as the backbone network of an autoencoder[19]. Kim et al. used Variational Auto Encoder to extract transient features and perform anonymous recognition[20]. NAITO et al. proposed a two-stage AE for feature extraction of detection parameters within a time window to achieve abnormality recognition[21]. However, the previous works done suffers from the following aspects, which prevent the large-scale application of data-based diagnosis methods using DL theory in production scenarios.

First, most data-based methods are fragile in the presence of disturbances. Due to the presence of electromagnetic interference in production environments and the absence of monitoring data due to failure of measurement points with initiating events, these disturbances can be catastrophically detrimental to the model making sound judgments. However, most of the existing works just consider noise interference only when validating the model. In reactor fault diagnosis, hardly any work considers the tolerance of the model to different levels of noise and crippled data at model construction.

Second, the validity of features extracted by DL models cannot be verified. In regression and classification tasks using DL, the first several layers of the network can be considered as feature extractors of the input data and the output of the network is obtained through the last layers of nonlinear mapping. However, in DL algorithms used for fault diagnosis, most of them use end-to-end training methods, so it is difficult to judge the validity of features.

Third, data-based diagnostic models are commonly considered as black boxes that do not provide operators with valid information for diagnostic purposes. Although the data-based approach can obtain a result with high accuracy, the internal reasoning process of the model cannot be understood by humans, so the operator still needs to re-perform manual diagnosis through a completely empirical-dependent manner.

Finally, each sampling point of the parameters has a significant correlation on the time axis, yet most models cannot combine the temporal attributes of the parameters with the parallel computing strength of the GPU. Most of the network models used for diagnosis use convolutional neural networks (CNN) or recurrent neural networks (RNN) as the backbone: 1) RNN-based networks can preserve the temporal characteristics of the data, but the algorithms cannot be parallelized due to their serial computing nature; 2) CNN-based networks cannot retain the temporal features of the original data, although they can be fully parallelized.

Therefore, there is a need for a more robust, feature-valid, interpretable diagnostic model that is also highly parallelable and preserves the temporal characteristics of the original data.

**1.4 Goal of our work**

In recent years, the flourishing of self-supervised representational learning techniques and interpretable techniques in computer science have provided inspiration for implementing desired diagnostic models.

In IFD of safety required scenarios, it is necessary to focus not only on the correctness of the diagnosis, but also on the robustness of the diagnosis process to disturbed monitoring data. Since deep learning techniques are a key implementation of data-based IFD and as deep learning mechanically derives abstract features of the input data automatically through multiple layers of nonlinear activation and mapping relationships, attention to the validity of the features is essential. Applying representation learning to IFD is a valuable research direction due to the fact that effective representations not only facilitate downstream diagnostic tasks but also help to migrate to authentic scenarios. Wang et al [22] extracted low-level features through a modified ResNet-50 network and then constructed a multi-scale feature learner to analyze high-level features as input to a fault classifier, solving the problem of incompatibility within the data domain of the training and test sets. Xia et al [23] proposed a digital twin (DT) and deep transfer learning-based framework for intelligent fault diagnosis of machinery, using a DT-based pre-trained sparse denoising autoencoder fine-tuned on the test domain to achieve a diagnosis task with few samples in real scenes. In IFD for nuclear reactor anomalous transients, Yang et al [24] constructed a diagnostic model based on LSTM and autoencoder and has the ability to identify untrained transient type features, i.e., the diagnostic model has the ability to say "I don't know".

Due to the growing concern for security, accountability, and auditing in the PHM, a growing number of researchers are applying eXplainable Artificial Intelligence (XAI) theories to counter the lesser application in engineering scenarios due to the "black box" nature of data-based models[25–30]. For decades, IFD technology has been focused more on diagnostic accuracy than diagnostic reliability, and XAI technology for IFD is a looming area of research to advance the Technology Readiness Level (TRL) of IFD technology[31].

Based on the above insight, to build an IFD system that can be used in production scenarios, this paper has conducted research related to the construction of the diagnostic model and interpreting the model, taking the loss of coolant accident (LOCA) of HPR1000 as the object. The major innovations are as follows:

First, Denoising Padded Autoencoder (DPAE), a robust feature extraction model for monitoring parameters with noise and missing data at the model structure level, is constructed. On the one hand, this model full uses the parallelism of the GPU while preserving the sequential features of the original data. On the other hand, the encoder-decoder structure enables the reconstruction of noise-added and randomly padded data to smooth and complete original data.

Second, a hierarchical initiating event diagnosis process without prior knowledge is proposed. This process first extracts valid low-dimensional representations of the monitored parameters using the encoder part of DPAE, and then uses the representation vectors as input data for arbitrary downstream diagnostic methods. It is experimentally verified that this hierarchical diagnosis method outperforms the end-to-end diagnosis method in terms of diagnostic accuracy for disturbed data.

Finally, an interpretable framework for diagnosis based on saliency transfer is designed. This interpretive approach not only provides the manipulator with a basis for making judgments by the IFD system, but is also agnostic to the downstream models in the hierarchical diagnostic process.

This paper is organized as follows. First, the proposed algorithm will be described in Section 2. In the third part, we will elaborate on the details of the LOCA diagnostic experiments for HPR1000 and the analysis of the experimental results. Finally, we will conclude the work and give an outlook.

## 2. Method

In this section, 2.1 will describe IFD for reactor initiation events in this paper, then the encoder part and decoder structure of the autoencoder for representation learning will be presented in 2.2 and 2.3, respectively, and finally the interpretation method of the diagnostic model will be described in 2.4.

### 2.1 Task description

The IFD process comprises three parts, which are data collection, feature extraction and status confirmation[32]. The proposed diagnostic method follows these three steps.

During data collection, 38 parameters are sampled at a frequency of twice per second for 100 seconds after the initiating event. Each parameter is represented by vectors as $u_d$, which is a sequence of samples on a time scale for the simulated signal $u(t)$. The selected detection parameters are concatenated to form the original tensor of two dimensions describing the initiating event $\mathcal{X}$. They are related as follows

$$\begin{aligned}
u_d(t) &= \sum_{n=1}^{\infty} u(nT)\delta(t-nT) \\
\boldsymbol{u}_d(n) &= u_d(nT) \in \mathbb{R}^{p \times 1}, n \in \{1,2,\ldots,p\} \\
\mathcal{X} &= \left[\boldsymbol{u}_{d,1}; \boldsymbol{u}_{d,2}; \ldots; \boldsymbol{u}_{d,l}\right] \in \mathbb{R}^{p \times l}
\end{aligned} \quad (1)$$

where $T$ is the sampling period; $p$ is the number of samples; and $l$ is the number of monitoring parameters.

In the feature extraction step, the encoder part of the DPAE is used to perform randomly masked feature extraction of the data collected by the monitoring system to obtain a low-dimensional representation vector. This process can be represented by a mapping, as

$$f_{\text{enc}} : \mathcal{X} \to \textbf{\textit{latent}} \in \mathbb{R}^d \tag{2}$$

where $f_{\text{enc}}$ is the encoder mapping; $\textbf{\textit{latent}} \in \mathbb{R}^d$ is the representation vector of the monitoring data; $d$ is the dimension of the representation vector and satisfies $d \ll p \times l$.

In the state confirmation step, the obtained representation vector is used as a proxy for the transient as an input parameter for the downstream state confirmation classifier or regressor. The mapping relationship is as follows

$$\begin{aligned} g_{\text{cla}} &: \textbf{\textit{latent}} \to \textbf{\textit{label}}_{\text{cla}} \\ g_{\text{reg}} &: \textbf{\textit{latent}} \to \textbf{\textit{label}}_{\text{sca}} \end{aligned} \tag{3}$$

where $\textbf{\textit{label}}_{\text{cla}}$ and $\textbf{\textit{label}}_{\text{sca}}$ are categorical and scalar diagnostic labels, respectively; $g_{\text{cla}}$ and $g_{\text{reg}}$ are mappings from representation vectors to categorical and scalar diagnostic labels, respectively.

## 2.2 Encoder of DPAE

There are three main functional modules in encoder, which are: 1) Pre-processing of raw monitoring data. In this process, it will insert noise and paddings, patchify the input data and insert learnable position encoding; 2) Transformer module based on multi-layer multi-headed self-attention mechanism. During this stage, the model will automatically capture the associations between the input data and inject representations about the monitored information into the class token; 3) A two-layer LSTM module for representational encoding. This will integrate all the associated features extracted after the Transformer module and focus on the representation information stored in the class token.

### 2.2.1 Pre-processing of raw monitoring data

First, the original monitoring data are supplemented with noise of different signal-to-noise ratios (SNR) to form a noise-laden monitoring data tensor $\mathcal{X}_{\text{n}} = \text{Noise}(\mathcal{X}; snr)$. This step allows the model to be exposed to noise during the training stage, so that the trained model itself is resistant to different levels of noise. Then, the monitoring data are patchified at the parameter level, as each patch contains a sequence of samples of the monitored parameters over a small-time range and can be considered as a patch that reflects the local high frequency information of the monitoring data, as

$$\begin{aligned} \left[ \textbf{\textit{patch}}_{i,1}; \textbf{\textit{patch}}_{i,2}; \ldots; \textbf{\textit{patch}}_{i,m} \right] &= \textbf{\textit{u}}_{\text{d},i} \\ \textbf{\textit{patch}}_{i,m} &\in \mathbb{R}^{(p/m) \times 1} \end{aligned} \tag{4}$$

where $i$ indicates the serial number of the monitoring parameter; $m$ indicates the number of patches a parameter has been patchified. Therefore, the noisy monitoring data $\mathcal{X}_n$ is transformed into a sequence of patches with patch size as the patch dimension $\mathcal{X}_p$, as

$$\begin{aligned}\mathcal{X}_p &= \sum_{i=1}^{l}\sum_{j=1}^{m} e_i \otimes e_j \otimes \textbf{\textit{patch}}_{i,j}^{\top} \\ &= \begin{bmatrix} [\textbf{\textit{patch}}_{1,1}; \textbf{\textit{patch}}_{1,2}; \cdots; \textbf{\textit{patch}}_{1,m}]^{\top} \\ [\textbf{\textit{patch}}_{2,1}; \textbf{\textit{patch}}_{2,2}; \cdots; \textbf{\textit{patch}}_{2,m}]^{\top} \\ \cdots \quad \cdots \quad \cdots \\ [\textbf{\textit{patch}}_{l,1}; \textbf{\textit{patch}}_{l,2}; \cdots; \textbf{\textit{patch}}_{l,m}]^{\top} \end{bmatrix} \\ &\in \mathbb{R}^{(l \cdot m) \times (p/m)} \xrightarrow[D=p/m]{N=l \cdot m} \mathbb{R}^{N \times D}\end{aligned} \quad (5)$$

where $e_i$ denotes the unit vector with the $i$-th element being 1; $\otimes$ is the Kronecker product operator. Next, the patch sequences were set to zero by a certain percentage to achieve model support for the crippled monitoring data. The rationale is that the way the padded patches are reflected in the original monitoring parameters is consistent with the failure mode of the monitoring points, i.e., into segments of time, rather than individual moments.

Finally, to automatically learn a valid representation of the input data during encoding, a learnable class token $x_{\text{class}} \in \mathbb{R}^{1 \times D}$ is added to the header of the patch sequence. Subsequently, since the patches reflects the low-frequency features in the monitoring parameters, which are supposed to be sensitive to location information, a learnable positional encoding of the same dimension $E_{\text{pos}} \in \mathbb{R}^{(N+1) \times D}$ is added to the patch sequence. After the above steps, the pre-processed monitoring data is obtained, as

$$\mathcal{X}_0 = \left[x_{\text{class}}; \mathcal{X}_p^{\top}\right]^{\top} + E_{\text{pos}} \in \mathbb{R}^{(N+1) \times D} \quad (6)$$

### 2.2.2 Transformer module for encoding

In the encoding stage, the patch sequence is first encoded by multiple Transformer blocks through a multi-layer and multi-headed self-attention mechanism, which automatically captures the remote correlation information between patches in a highly parallelized computation process. This computational process can be easily migrated to the GPU for acceleration. As the data passes through each Transformer module, it is encoded as

$$\begin{aligned}\mathcal{X}_q' &= \text{MSA}(\text{LN}(\mathcal{X}_{q-1})) + \mathcal{X}_{q-1} \\ \mathcal{X}_q &= \text{MLP}(\text{LN}(\mathcal{X}_q')) + \mathcal{X}_q', q \in \{1, 2, \cdots, Q\}\end{aligned} \quad (7)$$

where $Q$ is the number of Transformer blocks passed; MSA is the Multi-headed Self-Attention; LN is the Layer Normalization process; and MLP is the Multi-Layer Perceptron with a single hidden layer. MSA is the key aspect of the Transformer module, which is calculated as follows

$$[q, k, v] = \mathcal{X} U_{\text{qkv}}, \ \mathcal{X} \in \mathbb{R}^{N \times D}, \ U_{\text{qkv}} \in \mathbb{R}^{D \times 3D_h}$$

$$A = \text{softmax}\left(qk^\top / \sqrt{D_h}\right)$$

$$SA = Av \tag{8}$$

$$MSA = \left(\sum_{i=1}^{k} e_i \otimes SA_i(\mathcal{X})\right) U_{\text{msa}} \in \mathbb{R}^{N \times D}$$

where $D_h$ is the dimensionality of the patch after projection; and $U_{\text{msa}} \in \mathbb{R}^{(k \cdot D_h) \times D}$ is the multi-headed self-attention dimensionality transformation matrix. After encoding, the output patch sequence has the same dimension as the input patch sequence.

### 2.2.3 LSTM module for representational encoding

Although most of the representations of the monitoring data have been written into the class token, there may miss representations distributed in other patches, so it is necessary to traverse all patches including the class token. The traversal is done by a two-layer unidirectional Long Short-Term Memory (LSTM) recurrent neural network. The LSTM is characterized by a long memory time for the early input signals and the highest correlation of the internal state vector with the last input signal[33]. Therefore, if the class token is used as the last input, the state variables of the LSTM can incorporate all possible representational information, and the representational information in the class token is to be dominant. The LSTM module is calculated as follows:

$$\begin{cases} i_t = \sigma\left(W_{ii} \textbf{\textit{patch}}_t + b_{ii} + W_{hi} h_{t-1} + b_{hi}\right) \\ f_t = \sigma\left(W_{if} \textbf{\textit{patch}}_t + b_{if} + W_{hf} h_{t-1} + b_{hf}\right) \\ g_t = \tanh\left(W_{ig} \textbf{\textit{patch}}_t + b_{ig} + W_{hg} h_{t-1} + b_{hg}\right) \\ O_t = \sigma\left(W_{io} \textbf{\textit{patch}}_t + b_{io} + W_{ho} h_{t-1} + b_{ho}\right) \\ c_t = f_t \odot c_{t-1} + i_t \odot g_t \\ h_t = O_t \odot \tanh(c_t) \\ t \in \{1, 2, \cdots, N\} \end{cases} \tag{9}$$

where $N$ is the length of this vector sequence; $W, b$ are the weights and biases to be learned; and $\textbf{\textit{patch}}_N = x_{\text{class}}$. Since all the representation information is already contained in the state vector $c_N$, a low-dimensional state representation is obtained by a multi-layer perceptron with three fully connected layers $\textbf{\textit{latent}} = \text{MLP}(c_N)$. Thus, the work flow diagram of the whole encoder is shown in Figure 1.

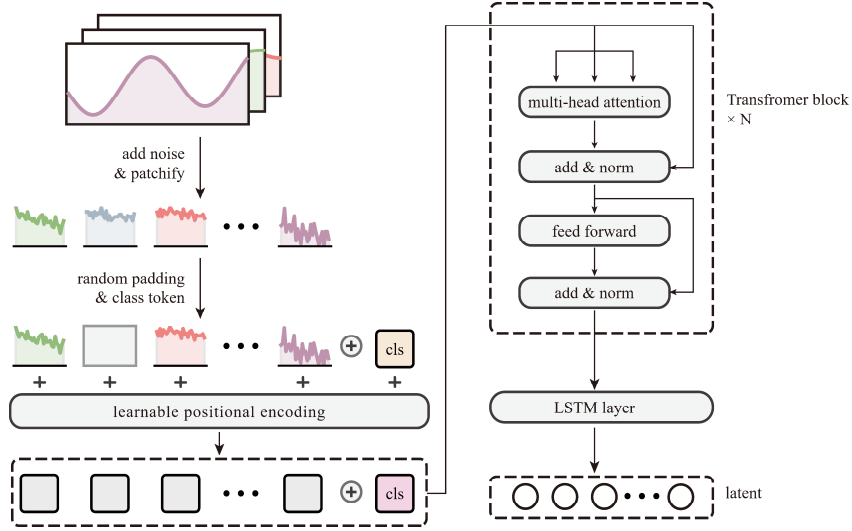

Figure 1 The block diagram of DPAE encoder

## 2.3 Decoder of DPAE

In the decoder part, there are three main steps, which are 1) Dimensional transformation and concatenation of class token. The purpose of this step is to transform the latent into a data shape that can be fed into the decoder, while preserving the category information of the original monitoring parameters, so that sufficient valid information about the incident anomaly transients can be used in the decoding process; 2) Adding a fixed positional encoding. The intention of this step is to preserve the relative positions of the elements in an explicit way during decoding; 3) Decoding and unpatchifying. After this step, the latent variables extracted from the transients containing noise and missing data are decoded into smooth and complete raw monitoring data.

### 2.3.1 Dimensional transformation and concatenation of class token

First, to facilitate the use of the same Transformer module employed by the encoder for decoding, the state representation vector is transformed to the same dimension as the patch sequence by a fully connected layer, i.e. $\mathcal{X}'_{de,0} = \mathrm{MLP}(latent) \in \mathbb{R}^{N \times D}$. Subsequently, the encoded class token from the encoder containing the class information is concatenated to form the input patch sequence for the decoder $\mathcal{X}_{de,0} = \left[ \boldsymbol{x}_{class}; \mathcal{X}'_{de,0} \right] \in \mathbb{R}^{(N+1) \times D}$.

### 2.3.2 Fixed positional encoding

To keep sufficient positional information during decoding, the patch sequence is added with a fixed positional encoding. The sin-cos encoding approach from [34] is used as

$$\begin{aligned} \boldsymbol{PE} &= PE_{ij}\boldsymbol{e}_{ij} \in \mathbb{R}^{(N+1) \times D} \\ PE_{pos,2m} &= \sin\left(pos / 10000^{2m/D}\right) \\ PE_{pos,2m+1} &= \cos\left(pos / 10000^{2m/D}\right) \end{aligned} \quad (10)$$

where $pos$ is the sequential number of the patch in the sequence; and $m \in \mathbb{N}_0$. Thus, the patch sequence to be put through the decoder is $\mathcal{X}_{de} = \mathcal{X}_{de,0} + \boldsymbol{PE} \in \mathbb{R}^{(N+1) \times D}$.

### 2.3.3 Decoding and unpatchifying

The dimensionality of the patch sequence has been restored to that of the encoder part and contains sufficient monitoring state information. To restore the implied state to the original monitoring parameter sampling sequence by the decoder part, multiple Transformer modules are still used here for decoding, and the process is shown in Eq.(7). After decoding, each region of the patch sequence contains local sampling information of the original monitoring parameters. Finally, the original noise-free and complete monitoring parameters are obtained by unpatchifying, as

$$\mathcal{X}_{re} = \left[ \boldsymbol{u}_{d,1}^{re}; \boldsymbol{u}_{d,2}^{re}; \ldots; \boldsymbol{u}_{d,l}^{re} \right] \in \mathbb{R}^{p \times l} \tag{11}$$

where $\boldsymbol{u}_{d,l}^{re}$ is the reconstructed sampling sequence of the $l$-th parameter. Thus, the diagram of the decoder part is shown in Figure 2.

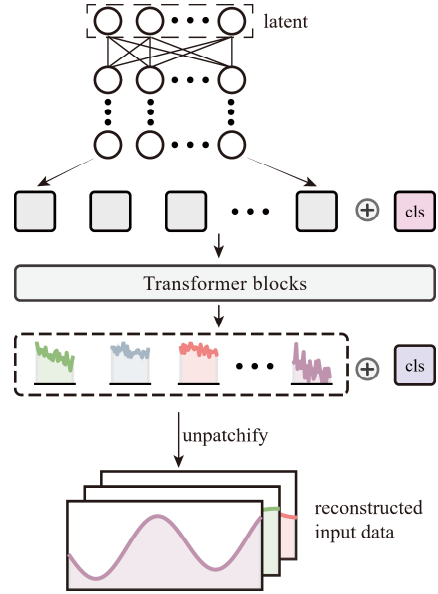

Figure 2 The block diagram of DPAE decoder

Therefore, after connecting the encoder and decoder, the overall structure of DPAE is shown in Figure 3.

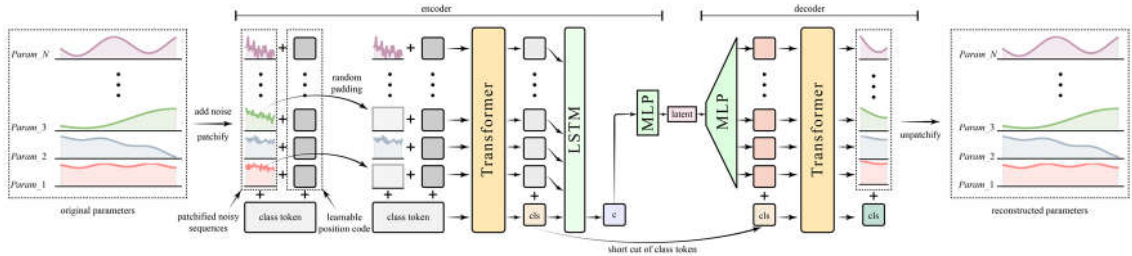

Figure 3 Overall structure of DPAE

### 2.4 Parameter importance interpretation method

To obtain the significance of each parameter on the diagnostic results, we adopt a post-hoc interpretable approach. Since the diagnostic approach proposed in this paper is under the classical IFD process, i.e., the data are first passed through an autoencoder to obtain a valid representation *latent*, and then the representation is transmitted as input to a downstream classifier $g_{cla}$ or regressor $g_{reg}$, rather than an end-to-end diagnostic process. A step-by-step approach to significance calculation is adopted. First, a well-trained downstream diagnostic model $g_{cla}$ or $g_{reg}$ is used to inferentially characterize the importance of each element in *latent* as $\boldsymbol{w} = w_j \boldsymbol{e}_j \in \mathbb{R}^d$. Then, the contribution of each parameter to each representation vector element is back-propagated through the well-trained encoder as $\boldsymbol{\Phi} = \varphi_{ij} \boldsymbol{e}_i \otimes \boldsymbol{e}_j \in \mathbb{R}^{l \times d}$, where $\varphi_{ij}$ represents the contribution of the parameter with index *i* to the representation element with index *j*.

## 3. Experiments and Discussion

In this section, the data sources used for model training will be described in 3.1; in 3.2, the ability of the model to extract representations will be validated in terms of data reconstruction and joint distribution of representation vectors; 3.3 will compare the diagnostic ability of different diagnostic methods on disturbed datasets; and finally, in 3.4, the importance of the elements of the representation vector and the input monitoring parameters will be analyzed by using the idea of importance cascade transfer with the post-hoc explainability of deep learning.

### 3.1 Dataset

The IFD process proposed in this paper uses the LOCA of the HPR1000 as the object of analysis, which is one of the advanced Gen III nuclear power plant models[35]. Once LOCA occurs, it could lead to core damage and possibly even a radioactive release. Therefore, if the LOCA break location and size can be determined quickly, robustly and credibly after an accident, it can provide a critical guide for effective subsequent disposal.

Since the diagnosis method in this study is data-based, reliable data sources are needed to ensure the quality and credibility of the model diagnosis. In the construction of the data-based diagnostic method for reactors, there are usually two sources of data experiments and simulation calculations, and the simulation-based data sources can be further divided into full-scale simulator software and system analysis programs. The data from experiments are scarce, difficult to obtain, and costly, so most of the data used for diagnostic model construction are from simulations.

The LOCA process is accompanied by a complex gas-liquid two-phase flow, which is manifested by significant pressure and flow pattern variations, and requires a simulation tool that is more refined for the wall heat transfer process. To ensure the accuracy of the simulation, the Advanced Reactor System Analysis Code (ARSAC) developed by Nuclear Power Institute of China (NPIC) was adopted as the simulation tool for the accident. ARSAC uses a non-equilibrium non-homogeneous Euler-Euler six equation two-fluid model that has undergone V&V and is therefore capable of capturing the complex thermodynamic phenomena expected under accident conditions[36]. In this study, the steady-state and transient input cards that can then be read by the ARSAC program are drawn according to the HPR1000 reactor system structure, and the system node diagram is shown in Figure 4. In the calculation of LOCA transients, the design parameters of the HPR1000 are used. the main technical parameters of the HPR1000 are shown in the Table 1.

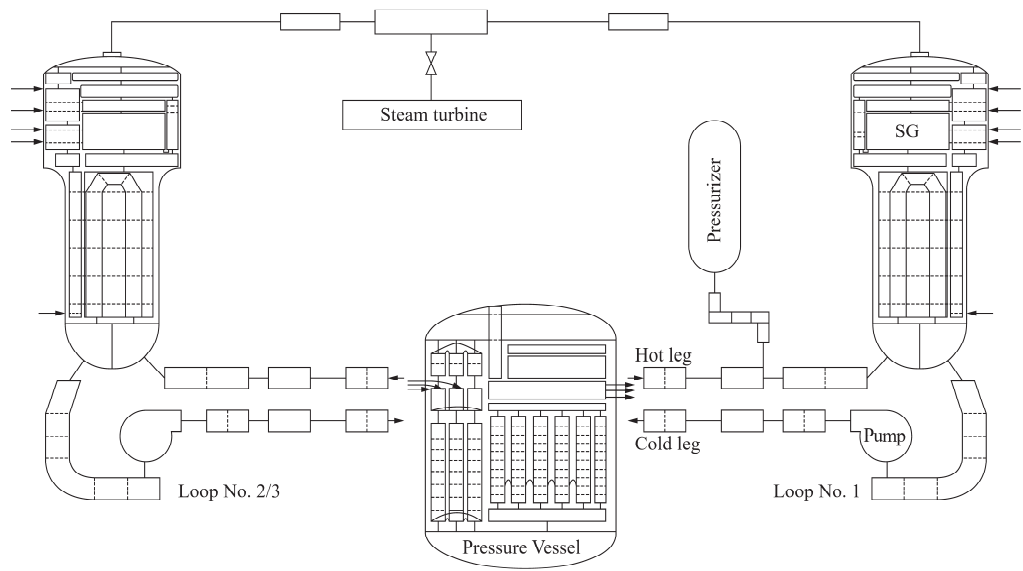

Figure 4 System node diagram of HPR1000

Table 1 Main technical parameters of the HPR1000[37]

| Parameter | Value |
| --- | --- |
| Reactor thermal output | 3050 MWth |
| Power plant output, gross | ~1170 MWe |
| Power plant output, net | ~1090 MWe |
| Power plant efficiency, net | ~36% |
| Mode of operation | Baseload and load follow |
| Plant design life | 60 years |
| Plant availability target | ≥90% |
| Refueling cycle | 18 months |
| Safety shutdown earthquake (SSE) | 0.3 g (g, gravitation constant) |
| Core damage frequency (CDF) | <$10^{-6}$ per reactor-year |
| Large release frequency (LRF) | <$10^{-7}$ per reactor-year |
| Occupational radiation exposure | <0.6 person-Sv per reactor-year |
| Operator non-intervention period | 0.5 h |
| Plant autonomy period | 72 h |

In the diagnostic task for LOCA, the most critical diagnostic aim is to identify the location and size of the break. Therefore, the data set was constructed with break locations containing cold and hot tube sections of coolant piping connected to the pressure vessel, and break sizes ranging from small breaks of 0.1 cm to large breaks of 35.1 cm in diameter. Each initiating event was represented using 38 monitoring parameters with a monitoring duration of 100 seconds and a sampling frequency of twice per second. The sampling values of the monitoring parameters are physical quantities extracted from the system model component nodes, and the correspondence between the names of the sampling nodes and the physical meanings of each parameter are shown in Table 2. After simulation, 356 sets of initiating events are calculated. During the training of DPAE, this paper uses all the initial events for training, which is because perturbations are randomly added in the pre-processing step each time the data passes through the autoencoder, so during the testing of Denoising Padded Autoencoder, it is never possible to have the same input data as during training.

Table 2 Correspondence of node names and parameters

| Node names | Parameters |
| --- | --- |
| tempf_505010000 | Temperature of main feed water |
| mflowj_505010000 | Mass flow rate of main feed water |
| cntrlvar_11 | Water level of steam generator |
| mflowj_566010000 | Mass flow rate of auxiliary feed water |
| mflowj_537000000 | Mass flow rate of main steam |
| p_540010000 | Pressure of steam line |
| p_850010000 | Pressure of steam busbar |
| voidf_811010000 | Water level of SI |
| p_810010000 | Pressure of SI |
| mflowj_811010000 | Mass flow rate of LHSI pump |
| mflowj_806000000 | Mass flow rate of boron injection pump |
| rktpow | Avg. power |
| cntrlvar_100 | Maximum average temperature of loops |
| tempf_138010000 | Temperature of reactor core outlet |
| tempf_155010000 | Temperature of the upper head |
| cntrlvar_2 | Water level of pressure vessel |
| p_155010000 | Pressure of reactor coolant |
| p_260010000 | Pressure of pressurizer |
| cntrlvar_42 | Water level of pressurizer |
| cntrlvar_121 | Mass flow rate of reactor coolant |
| tempf_200010000 | Temperature of the broken loop (1#) hot leg |
| tempf_300010000 | Temperature of hot leg of loop 2# |
| tempf_400010000 | Temperature of hot leg of loop 3# |
| tempf_250010000 | Temperature of the broken loop (1#) cold leg |
| tempf_350010000 | Temperature of cold leg of loop 2# |
| tempf_450010000 | Temperature of cold leg of loop 3# |
| cntrlvar_101 | Avg. temperature of the broken loop (1#) |
| cntrlvar_102 | Avg. temperature of loop 2# |
| cntrlvar_103 | Avg. temperature of loop 3# |
| pmpvel_235 | Pump speed of the broken loop (1#) |
| pmpvel_335 | Pump speed of loop 2# |
| pmpvel_435 | Pump speed of loop 3# |
| tempf_2700(1-5)0000 | Temperature of pressurizer surge tube (divided into 5 nodes) |
| tempg_260010000 | Gas temperature of pressurizer |
| tempf_262010000 | Liquid temperature of pressurizer |
| tempg_281010000 | Upstream temperature of the safety valve of the pressurizer |
| voidf_200010000 | Water level in the hot leg of the breakout loop |

**3.2 Validity of representation**

*3.2.1 Experiment details*

During the construction of DPAE, the hyper-parameters of the network structure are set as follows in this paper. First, since each parameter has 200 samples in 100 seconds, the vector dimension of each patch block is set to $D=40$, so the length of the patch sequence is $N=190$. In the encoder part, the number of Transformer blocks is $depth_{enc}=4$, where the number of heads of the multi-headed self-attention mechanism is $heads_{enc}=4$. For the LSTM recurrent neural network used to traverse the encoded patch sequences, the number of network stacking layers is set to 2, and the dimension of the state vector is 40. In the decoder part, the number of Transformer blocks is $depth_{dec}=4$ and the number of heads of the multi-headed self-attention mechanism is $heads_{dec}=4$. The dimensionality scaling of the feed-forward neural network intermediate layer used in the network is $ratio_{mlp}=0.8$. The dropout regularization means with a scale of 0.1 is used.

To train the network, the optimization objective is the mean square error between the initial monitoring parameters $\mathcal{X}$ and the reconstructed parameters $\mathcal{X}_{re}$, as $loss = \mathrm{MSE}(\mathcal{X}, \mathcal{X}_{re}) = \frac{1}{N \times D} \|\mathcal{X}, \mathcal{X}_{re}\|_2$. The optimization algorithm of the network is the Adam algorithm with the addition of Nesterov Momentum, which has a more flexible, adaptive learning rate than the traditional gradient descent algorithm and can improve the phenomenon of falling into local minima during the optimization[38]. For the optimization algorithm, the initial learning rate is set to $lr_{ini}=1.0\times10^{-3}$, and the coefficients used to calculate the running average of the gradient and its square are $\beta_1=0.9$ and $\beta_2=0.999$. The smoothing coefficient is $\epsilon=1.0\times10^{-8}$, and the momentum decay coefficient is $4.0\times10^{-3}$.

In each training step for a sample, DPAE needs to reconstruct the data with different levels of noise and random masking ratios added, i.e., noise reduction and complementation. In the training procedure, the autoencoder first learns separately for data with high to low perturbation levels, and then repeats the learning several times for the perturbation level of focus. The advantages of this approach are: 1) Firstly, when learning the data with high perturbation level in the initial stage, the parameters to be learned are updated faster due to larger reconstruction error; 2) Secondly, when learning the data with less perturbation in the latter stage, the reconstruction error is smaller and the parameters are updated more slowly; 3) Finally, when learning the data with specific perturbation level for multiple repetitions, if this perturbation level is more consistent with the actual application scenario, the optimization process has a stronger focus on the application scene. In this paper, in each iteration step, the order of signal-to-noise ratio is $snr=[20.0,\ 35.0,\ 40.0,\ 30.0,\ 30.0]$, and the order of random masking ratio is $ratio_{pad} = [0.40, 0.25, 0.10, 0.20, 0.20]$. The training was iterated 1000 times on the full dataset.

### 3.2.2 Reconstruction of monitoring data

After the training of the DPAE is completed, the first thing that needs to be verified is whether this autoencoder can effectively complement and reduce the noise of the perturbed data, as this indicates that the encoder can perceive effective representation information for the abnormal transient. Although the overall reconstruction error of the monitoring parameters is small enough from the loss function, it still needs to be evaluated from the perspective of specific parameters. We select a LOCA initiation event with a break diameter of 15.1 cm and in the cold leg as the subject and plot a comparison of the original data, the data with noise and masking added, and the reconstructed data for some parameters with different padding and noise perturbations. The reconstruction data of the autoencoder when increasing the noise to $snr = 25.0$ and the padding ratio to $ratio_{pad}=0.40$ are considered for each parameter, based on $snr = 30.0$ and $ratio_{pad}=0.20$, respectively. Figure 5 to Figure 7 correspond to the trends of water level of the steam generator, mass flow rate of the main feed water and coolant pressure, respectively. It can be seen from these plots that the DPAE can reconstruct the monitoring data fairly well with the addition of different levels of perturbation.

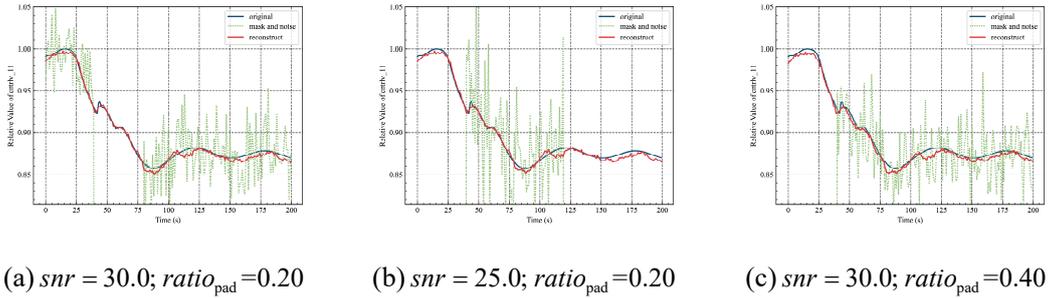

(a) $snr = 30.0$; $ratio_{pad}=0.20$     (b) $snr = 25.0$; $ratio_{pad}=0.20$     (c) $snr = 30.0$; $ratio_{pad}=0.40$

Figure 5 Reconstruction of steam generator water level

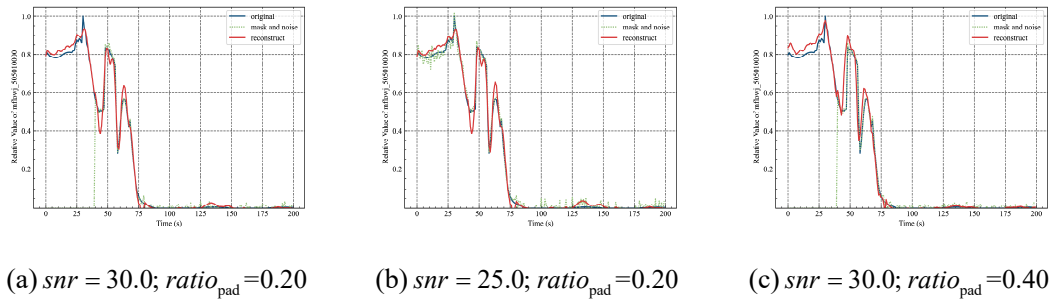

(a) $snr = 30.0$; $ratio_{pad}=0.20$     (b) $snr = 25.0$; $ratio_{pad}=0.20$     (c) $snr = 30.0$; $ratio_{pad}=0.40$

Figure 6 Reconstruction of the main feedwater mass flow rate

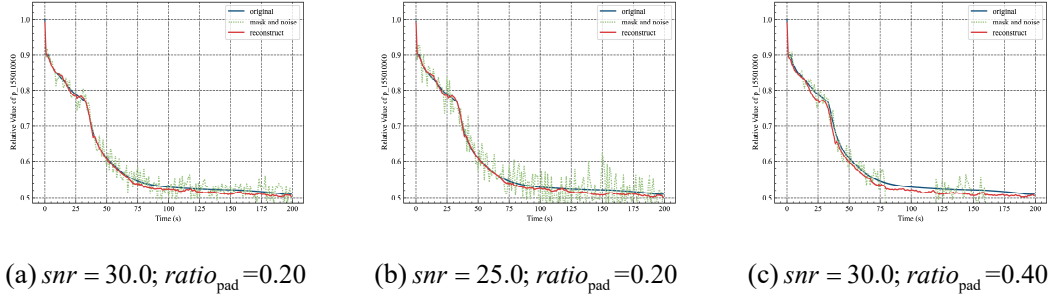

(a) $snr = 30.0$; $ratio_{pad} = 0.20$     (b) $snr = 25.0$; $ratio_{pad} = 0.20$     (c) $snr = 30.0$; $ratio_{pad} = 0.40$

Figure 7 Reconstruction of coolant pressure

*3.2.3 Validation of latent variables as transient representations*

Meanwhile, to ensure that the representations of the monitoring data extracted by the encoder are meaningful, i.e., the low-dimensional latent variables reflect the overall characteristics of the original monitoring data, they can be illustrated by the probability density distribution of the samples. In the unpreprocessed sample data, due to the similarity between the transient process of $\mathcal{X}_i$ and $\mathcal{X}_j$ caused by the initiating event, the joint probability can be defined by the similarity between the samples as $p_{ij}$, and then the probability density distribution in the sample space can be constructed as $P$. To visualize the distribution in 2D space, the t-distributed stochastic neighbor embedding (t-SNE) method is used to construct samples in 2D space $\boldsymbol{y}_1,\ldots,\boldsymbol{y}_N$ (with $\boldsymbol{y}_i \in \mathbb{R}^2$), using the probability density distribution $Q$ in 2D space instead of the distribution of samples $P$ in higher dimensional space. First, the distribution visualization of the original unpreprocessed 356 samples is shown in Figure 8. The larger scatter size and darker color in the figure represent the breaks of larger sizes, with the red area being the break occurring in the hot leg and the blue area being the break in the cold leg. It can be seen in the figure that there is a significant cluster distribution feature on the raw data.

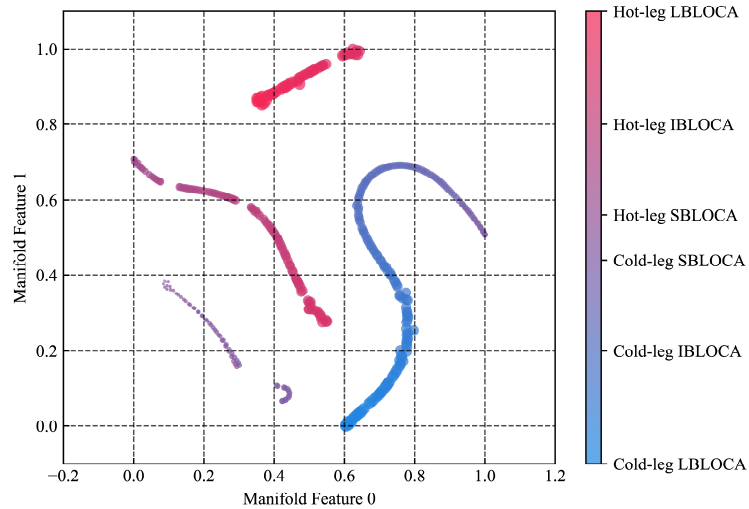

Figure 8 Distribution characteristics of the original samples

After pre-processing, a patch mask with a random ratio of $ratio_{pad}=0.30$ and noise with a signal-to-noise ratio $snr=30.0$ is added. Next, the distribution of these samples with added interference is visualized, as shown in Figure 9. As observed, by this time, the data has lost its meaningful distribution characteristics and has become a chaotic mess.

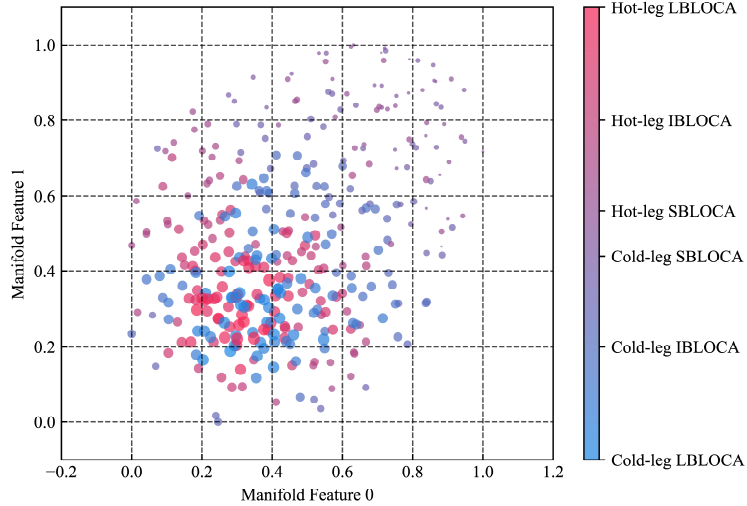

Figure 9 Characteristics of the sample distribution after adding interference

Next, the samples are fed into the trained DPAE encoder to obtain the representation vectors corresponding to these samples, as $latent_1,\cdots,latent_N$ (where $latent_i \in \mathbb{R}^D$). The visualization of the representation vectors is shown in Figure 10. As is clear from the figure, the representation vectors drawn from the disturbed samples have re-displayed meaningful distribution characteristics, and their cluster characteristics have obvious similarity with those of the original samples, i.e., they are roughly divided into five clusters. It is worth noting that the absolute value of the coordinates in the distribution feature map of the samples in two-dimensional space does not have a clear meaning, so attention should be paid to the relative positions between the sample points. Therefore, following the above analysis, it can be assumed that the sample features extracted by DPAE are meaningful and can replace the original sample data for downstream diagnostic tasks.

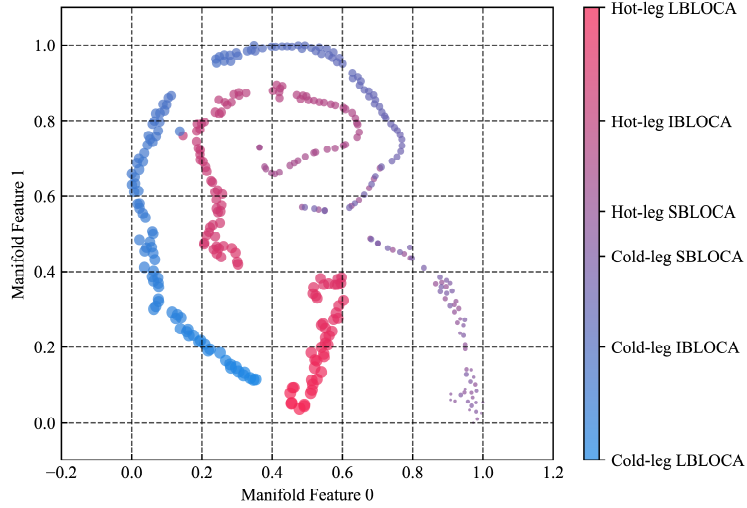

Figure 10 Distribution characteristics of the representations extracted from the disturbed samples

## 3.3 Capability of diagnosis

### 3.3.1 Two general types of diagnosis

The proposed diagnostic method is divided into two steps, "upstream" and "downstream", which firstly extracts valid representations of the disturbed monitoring parameters in the "upstream", and then the representation is provided to the "downstream" diagnostic model as a proxy for the sample. The diagnostic logic is shown in Figure 11.

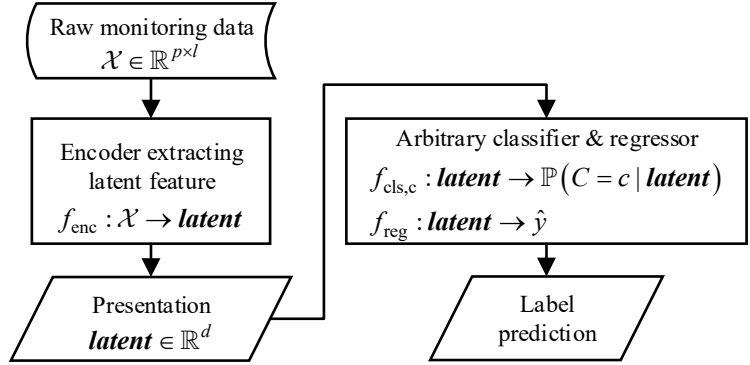

Figure 11 Flowchart of feature extraction first and then diagnosis

In contrast to the method proposed in this paper, there is an "end-to-end" diagnostic method in which the raw monitoring data, after noise addition and partial absence, is directly used as input to the diagnostic model. As shown in Figure 12.

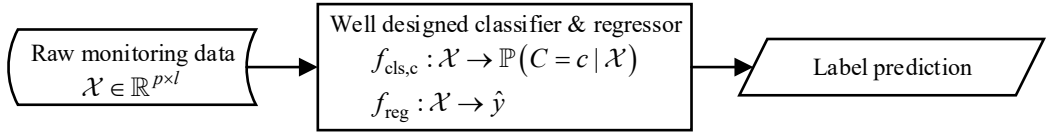

Figure 12 Flowchart of end-to-end diagnosis

For the "upstream" and "downstream" types of diagnostics, the performance of different shallow statistical learning-based classifiers and regressors will be compared, including random forests[39], SVM[40], XGBoost[41], and MLP. For the "end-to-end" type of diagnostics, the performance of previously proposed end-to-end classifiers and regressors, such as those based on BPNN[42] and CNN[12], as well as the author's own previously proposed TRES-CNN[43] approach, will be compared. The baseline models of these two types of diagnostic methods will be compared in an integrated manner.

To evaluate the different diagnostic methods, uniform evaluation criteria need to be determined. For the diagnosis of LOCA break location, the F1 value is selected as the evaluation index since it is a classification problem; for the diagnosis of LOCA break size, the root-mean-square error (RMSE) is selected as the evaluation index since it is a regression problem. The calculation method of the metric is shown below

$$F_1 = \frac{2}{recall^{-1} + precision^{-1}} \tag{12}$$

$$RMSE = \sqrt{\frac{\sum_{t=1}^{T}(\hat{y}_t - y_t)^2}{T}} \tag{13}$$

where $\hat{y}_t$ is the predicted break size; $y_t$ is the real break size label; and $T$ is the number of samples in the test dataset. To ensure the fairness of the experiment, the end-to-end model size is matched to the sum of the "upstream" plus "downstream" model sizes. When the objective function value on the validation set is updated by less than 1.0% in 20 iteration steps, the training of XGBoost and neural network-based models is early stopped, which is used not only to model regularization but also to ensure the conditional consistency of different models during the training process

*3.3.2 Comparison of diagnostic capability*

Each model uses the full sample of perturbed originating events as the training set for model learning. After training is completed, a separate random seed is set and based on this, new perturbed data is constructed for all samples. These new scrambled data are used as the test set. Since none of the samples in the test set had appeared in the training set, it was valid to use the newly generated perturbed monitoring data as the evaluation of the model. After the experiments on the test set, the performance results of each diagnostic method are shown in Table 3. As shown in the table, the diagnostic method with feature extraction followed by diagnosis has a significant improvement in performance compared to the end-to-end diagnostic method. This is attributed to the hierarchical diagnosis process, i.e., it is less difficult to train a model that assigns feature extraction and diagnosis to separate models than it is to train a larger model that performs both feature extraction and diagnosis tasks. Also, since the dimensionality of the representation vectors extracted from the disturbed monitoring data is sufficiently low, even using classical statistical learning models, such as SVM and XGBoost, can achieve performance close to or better than that of deep learning algorithms centered on connectionism. This means that with sufficiently robust pre-trained models for representation extraction, it is possible to build accident diagnosis models that are easy to maintain and tolerant of noise and local data residuals.

Table 3 Performance of different diagnostic methods and models

| Type | Method | Total model size (MB) | Cold leg precision | Hot leg precision | Macro-F1 | RMSE |
|---|---|---|---|---|---|---|
| Upstream & Downstream | MLP | 114.73 | 0.905 | 0.934 | 0.919 | **0.312** |
| | SVC/SVR | 114.47 | 0.898 | 0.917 | 0.908 | 0.649 |
| | XGBoost | 115.98 | **0.911** | **0.946** | **0.927** | 0.564 |
| | Random Forest | 115.64 | 0.858 | 0.871 | 0.864 | 0.982 |
| End to end | TRES-CNN | 116.21 | 0.682 | 0.646 | 0.662 | 2.945 |
| | BPNN | 115.32 | 0.612 | 0.588 | 0.598 | 3.623 |
| | CNN | 117.49 | 0.641 | 0.675 | 0.656 | 3.259 |

### 3.4 Explanation of Diagnostic Results

*3.4.1 Principles of explanation*

In safety-critical situations such as reactor fault diagnosis, the operator's decisions are directly or indirectly influenced by the operation assistance system, so the judgments made by this system need not only to be highly accurate but also to provide reasons that are convincing to the operator and thus assist the manual diagnosis. The scenario applies to the target audience of XAI as mentioned in [44]. The post-hoc interpretation of the model not only helps to understand the reasons for the model's judgments, but also serves the manipulator's judgment and decision-making process. This interpretation method is achieved by calculating the magnitude of the influence of each input parameter on the decision output. Since the diagnostic method proposed in this paper is a two-step process, it is necessary to calculate the importance of the elements in the representation vector first and then the importance of the input parameters.

Next, the Middle Break LOCA (MBLOCA) part of the operating conditions will be used to analyze which parameters contribute more significantly to the diagnostic results in the diagnostic process of this type of initiation event. The reason for choosing MBLOCA as the object of explanation lies in the fact that in the process of small and tiny breaks, the reactor core is not easily to be exposed under regulating the safety system; while in the process of large breaks, the water loss and pressure relief phenomena are significantly influenced by the location and size of the break, and the diagnostic process is simpler. It is noteworthy that, in the importance's analysis of the characterization vector elements, all samples were used; while in the analysis of the importance of the monitoring parameters, only the samples within the MBLOCA interval were used. This is because the contribution assignment of the characterization vector elements is global, while the parameter importance assignment is highly correlated according to the specific initiating event and is therefore local.

*3.4.2 Importance of representation vector elements*

Since the dimensionality of the representation vector used as input data in the downstream diagnostic task is small, only 128 dimensions, the SHapley Additive exPlanations (SHAP) based interpretation method with high consistency is used[45]. Since SHAP gives the principle of linear additivity of Shapley values, the input element contribution is uniquely determined in a single inference process. SHAP method is based on game theory and calculates the contribution of each participant of the game, to the outcome of the game, and the metric of contribution is the Shapley value[46]. In this paper, the participants of the game are each element of the representation vector, the process of the game is the mapping from the representation vector to the data labels, and the game outcome is the sample labels that the model needs to infer.

In the process of downstream diagnosis using representation vectors, considering the simplicity of the representation, the input parameters are representation vectors $x = latent = x_i e_i \in \mathbb{R}^D$, with which a binary vector is paired as $z = |\text{sgn}(x_i)| e_i \in \{0,1\}^D$. The inference mapping for the diagnosis is $g(x)$, defined as shown in Eq.(3). The XGBoost classifier and regressor trained in the previous section are used here. The explanation model $p(z)$ is defined as $p(z) = \phi_0 + \sum_{i=1}^{D} \phi_i z_i$, where $\phi_i$ is the attribution of the input parameters $x_i$ to the output, i.e., the Shapley value, which can be calculated by the following equation

$$\phi_i = \begin{cases} E_x[g(x)] & i = 0 \\ z_i E_{x \setminus x_i}[g(x) | x_i] & i = 1 \\ z_i \begin{cases} E_{x \setminus \{x_1, \cdots, x_i\}}[g(x) | \{x_1, \cdots, x_i\}] \\ -E_{x \setminus \{x_1, \cdots, x_{i-1}\}}[g(x) | \{x_1, \cdots, x_{i-1}\}] \end{cases} & i \geq 2 \end{cases} \quad (14)$$

where $E_{x \setminus \{x_1, \cdots, x_i\}}[g(x) | \{x_1, \cdots, x_i\}]$ denotes the expectation of $g(x)$ on the complementary set of representational elements $x \setminus \{x_1, \cdots, x_i\}$ given the set of representational elements $\{x_1, \cdots, x_i\}$. The expectations of each individual condition are approximated by the model independent Kernel SHAP method[45]. After statistical analysis, the Shapley values calculated for the representation vector elements numbered $i$ on the $j$-th sample in the classification and regression tasks are $\phi_{ij}^{\text{cla}}$ and $\phi_{ij}^{\text{reg}}$, respectively. The summary plots for the classification and regression tasks are shown in Figure 13(a) and Figure 13(b), respectively, and only the top 20 elements in terms of importance are shown in the plots.

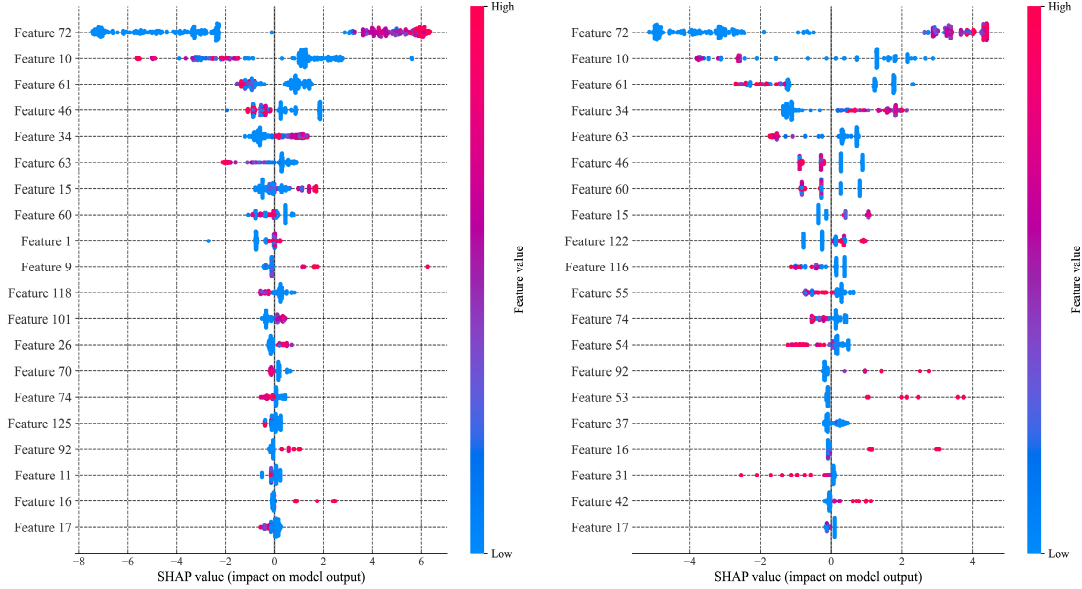

(a) Classification tasks (break location)   (b) Regression tasks (break size)

Figure 13 Shapley value distribution of the elements of the characterization vector's influence on the diagnostic output

Then the absolute values of the Shapley values corresponding to each representational element in the two types of tasks are summed up $\varphi_{ij} = \left|\phi_{ij}^{\text{cla}}\right| + \left|\phi_{ij}^{\text{reg}}\right|$, and finally the importance of each representational element is obtained as

$$\boldsymbol{\varphi} = \varphi_i = \frac{\sum_j \varphi_{ij}}{\sum_i \sum_j \varphi_{ij}} \boldsymbol{e}_i \in \mathbb{R}^D \quad (15)$$

The importance and ranking of the representation vector elements are shown in the Figure 14. As with the summary plots, the ranking only shows the top 20.

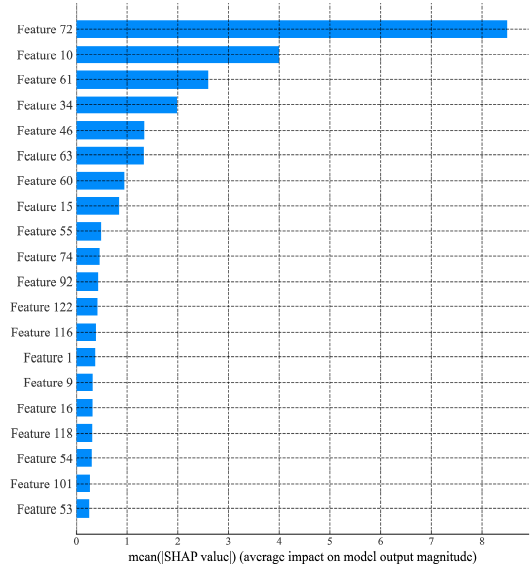

Figure 14 Importance ranking of representation vector elements

### 3.4.3 Importance of monitoring parameters

In the importance analysis of the monitoring parameters, a feature ablation-based calculation method is used. The principle of this method is to use a baseline to replace each input feature one after another and to calculate the difference in the output of the DPAE encoder part before and after the replacement. The baseline for ablation is set to 0.5 due to the encoder's ability to tolerate local paddings, i.e., a padding of 0 may not cause a significant change in the representation vector. in addition, the region of each feature ablation is half a patch size, i.e., a vector of dimension 20, used to distinguish it from the random masking process in preprocessing

The reason for not choosing to compute the Shapley values anymore is that computational cost is too large in terms of calculation complexity. Since interpreting the encoder can only be performed using the model-independent Kernel SHAP method, which requires perturbing the input parameters near each sample point and training a simple proxy model consequently. A large computational overhead is required for cases where the input parameters are too large, i.e., 7600 sample points in each monitoring data. And the noise tolerance of the encoder leads to the fact that a small perturbation of the monitoring data does not significantly change the output of the representation vector.

In an abnormal transient, let the contribution of the $n$-th ablation region of the parameter numbered $m$ to the $i$-th characterization vector element of the encoder output be $\omega_{mn}^i$. Then the contribution of the parameter numbered $m$ in this condition is calculated as

$$\psi_m = \sum_n \sum_i \omega_{mn}^i \varphi_i \qquad (16)$$

After the importance analysis of the input parameters for eight initiating event conditions containing cold leg and hot leg breaks with size in $[9.1\text{cm}, 9.7\text{cm}]$, a heat map of the importance distribution $\omega_{mn} = \sum_i \omega_{mn}^i \varphi_i$ was drawn, as shown in Figure 15. Then, the importance of different feature ablation regions of each parameter was accumulated and ranked to obtain the importance ranking of each monitoring parameter in the above working condition interval, as shown in Figure 16

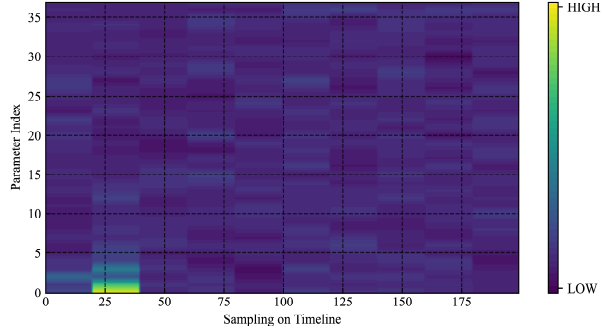

Figure 15 Heat map of importance distribution of each ablation area

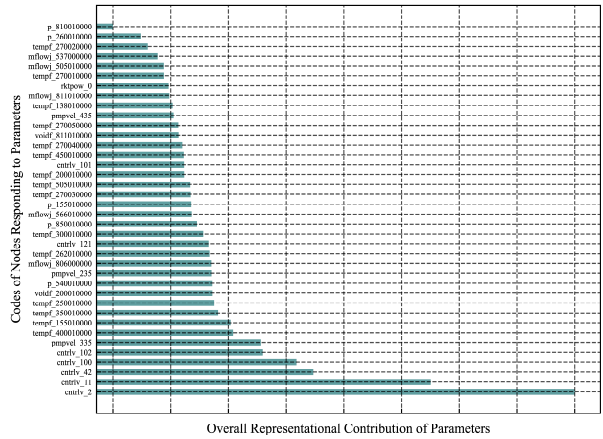

Figure 16 Contribution of input parameters to diagnosis between experimental conditions

The basic logic of diagnosis is that the changes in the parameters of the initiating event lead to corresponding system response changes, and the diagnostic model can invert the parameters of the initiating event by the variability of the system response. Therefore, only by indicating that there will be significant differences of important parameters under diverse initiating conditions, can the interpretation of the model make sense. To demonstrate this, the normalized response of the top 5 parameters in importance were extracted, where the highest-ranking indexes start from 0, as shown in Figure 17 to Figure 21. As noted from the figures, the trend of the parameters ranked ahead in importance can reflect the characteristics of different working conditions of the initiating event, so the diagnosis method proposed in this paper is effective and conforms to the basic logic of diagnosis. Meanwhile, by extracting the parameters of interest to the diagnostic model, it can be used as auxiliary information for the operator's manual diagnosis. It allows the operator to know not only the result of the diagnosis but also the reason for the judgment made by the diagnosis model.

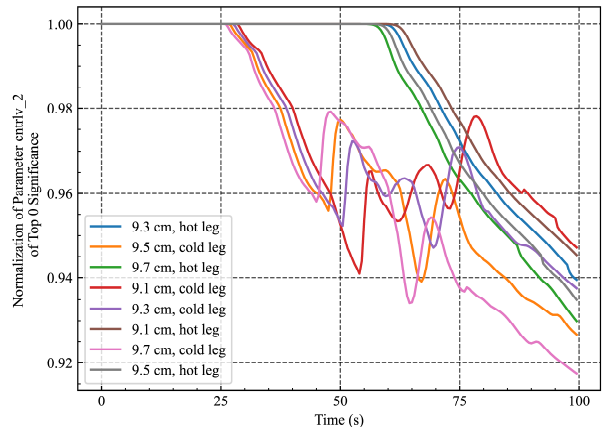

Figure 17 Top 1 parameter in terms of importance (pressure vessel water level)

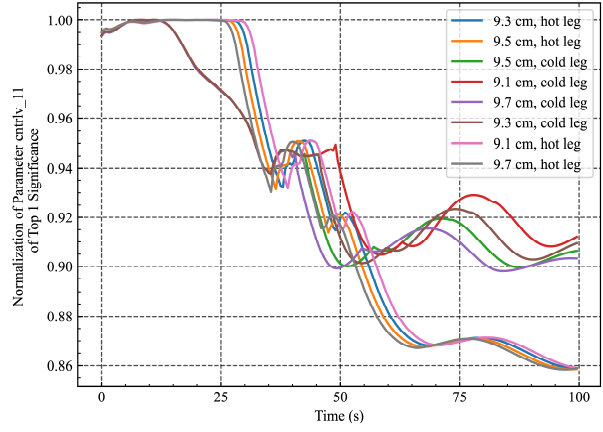

Figure 18 2nd most important parameter (steam generator water level)

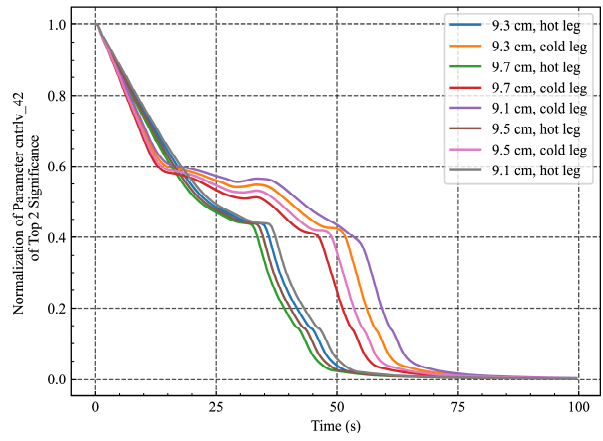

Figure 19 3rd most important parameter (pressurizer water level)

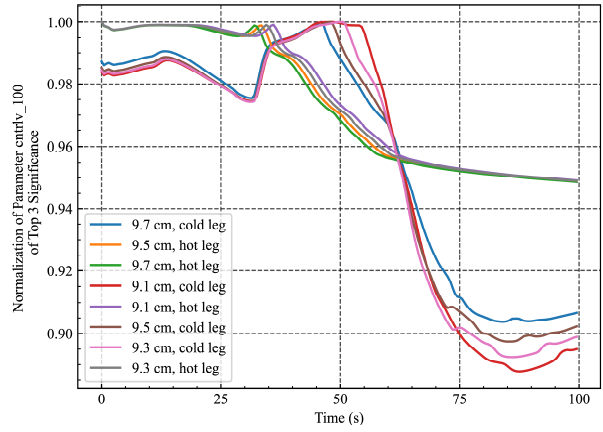

Figure 20 4th most important parameter (maximum average temperature of the loop)

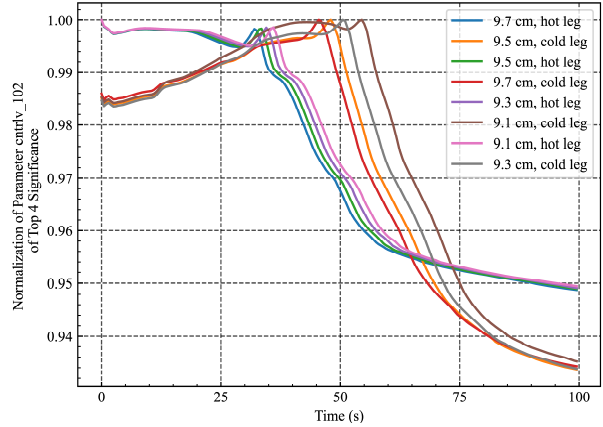

Figure 21 5th most important parameter (2# loop average temperature)

# 4. Conclusions

This paper focuses on the IFD process of reactor systems and proposes a novel robust and interpretable diagnostic framework. The main work is divided into three points: 1) First, a novel autoencoder Denoising Padded Autoencoder is proposed for representation extraction of monitoring data, and this representation extractor is still effective for applications with disturbed data with signal-to-noise ratio up to 25.0 and monitoring data missing up to 40.0%; 2) Second, a diagnostic framework is proposed using Denoising Padded Autoencoder's encoder part to extract the representations, and then a shallow statistical learning algorithm for diagnosis is proposed as a diagnostic framework. As tested on a disturbed dataset, this stepwise diagnostic approach has 41.8% and 80.8% higher classification and regression task evaluation metrics, respectively, compared to the end-to-end diagnostic approach; 3) Finally, a hierarchical interpretation algorithm using SHAP and feature ablation is proposed to analyze the importance of the input monitoring parameters and to validate the effectiveness of the high importance parameters. The results provide a reference method for constructing a robust and interpretable intelligent reactor anomaly diagnosis system

# Credit author statement

**Chengyuan Li**: Conceptualization, Methodology, Investigation, Software, Writing-original draft. **Zhifang Qiu**: Writing-Review & Editing. **Zhangrui Yan**: Visualization, Validation, Writing-Review & Editing. **Meifu Li**: Supervision, Data Curation, Formal analysis.

# Declaration of competing interest

The authors declare that they have no known competing financial interests or personal relationships that could have appeared to influence the work reported in this paper.